\documentclass{article}

\usepackage{amsmath}
\usepackage{hepunits}
\usepackage{graphicx}
\usepackage{relsize}
\usepackage{hyperref}

\renewcommand{\d}{\text{d}}

\begin{document}
\newcommand{\ind}[1]{${}^{\textnormal{#1}}$}
\title{Dynamic Alignment Method at the LHC}
\author{%
R. Staszewski\ind{a,}\footnote{e-mail: \texttt{rafal.staszewski@ifj.edu.pl},}\ ,\ \ 
M. Trzebi\'nski\ind{a,}\footnote{e-mail: \texttt{maciej.trzebinski@ifj.edu.pl}}\ ,\ \ 
J. Chwastowski\ind{b,a,}\footnote{e-mail: \texttt{janusz.chwastowski@ifj.edu.pl}}
\\[5mm]
\ind{a}%
\small H. Niewodnicza\'nski Institute of Nuclear Physics\\[-3mm]
\small Polish Academy of Sciences\\[-3mm]
\small ul. Radzikowskiego 152, 31-342 Krak\'ow, Poland.\\[5mm]
\ind{b}%
\small Institute of Teleinformatics\\[-3mm]
\small Faculty of Physics, Mathematics and Computer Science\\[-3mm]
\small Cracow University of Technology\\[-3mm]
\small ul. Warszawska 24, 31-115 Krak\'ow, Poland\\[5mm]
}
\maketitle

\begin{abstract} The dynamic alignment method of the forward proton detectors
proposed by the CDF Collaboration is reviewed. Applicability of the method at
the LHC is discussed. \end{abstract} 

\section{Introduction}

A typical $pp$ collision at the LHC consists of several parton-parton
interactions. This causes the color charge flow between the protons and leads
to color dipoles creation. In consequence, the dipoles radiate filling the
detector with particles. However, in a fraction of events the protons interact
coherently, either electromagnetically -- exchanging a photon, or strongly --
via an exchange of a color singlet object named Pomeron. In such collision one
or both protons, staying intact, may lose some part of their energy and be
scattered at very small angles into the accelerator beam pipe. Since the
exchanged object is a color singlet, a suppression of particle radiation is
observed in such events. This leads to a presence of rapidity regions devoid of
particles -- the rapidity gaps.

When both protons stay intact and two emitted photons or Pomerons interact with
each other producing a state in the central part of the rapidity space, the
event is called a Central Exclusive Production (CEP) \cite{Albrow:2010yb}.
Various objects can be produced in such a process: particles ($\chi_c$, Higgs
boson, $J/\psi$) or systems of particles (these may: be a pair of leptons,
photons, jets, SUSY particles, \textit{etc.}). At hadron colliders such
processes give a unique opportunity to measure all final state particles, which
leads to kinematic constrains and results a good resolution of the centrally
produced system mass reconstruction in a wide range of masses
\cite{Staszewski:2009sw}. This can be done with proton tagging detectors that
are inserted into the beam pipe far away from the Interaction Point (typically
several dozens or even hundreds of metres).

At the LHC, both ATLAS and CMS Collaborations are equi\-pped with such
detectors -- ALFA (Absolute Luminosity For ATLAS) and TOTEM (Total Cross
Section, Elastic Scattering and Diffraction Dissociation), respectively.
However, both detector systems are designed to work during dedicated LHC runs
with special machine tune (the so-called high $\beta^*$ optics), when
the luminosity is a few orders of magnitude smaller than the nominal
\unit{$10^{34}$}{\lumiunits}. Therefore, both Collaborations plan to install
additional detectors that will be able to work in the standard LHC tune
environment: the AFP (ATLAS Forward Physics) detectors and the HPS (High
Precision Spectrometer) for ATLAS and CMS respectively.

Forward proton detectors can measure the position and direction of the
scattered proton trajectory.  From that measurement, the fractional momentum
loss, $\xi = (E_0 - E')/E_0$, and the four-momentum transfer, $t \approx p_T^2
= p_x^2 + p_y^2$, can be reconstructed (actually, the value of four-momentum
transfer is negative and $t$ denotes hereafter its absolute value).  A
crucial element for the reconstruction resolution is proper alignment of
detectors, \textit{i.e.} a precise knowledge of their position. One should note
that detectors position (their distance from the beam) has to be adjusted
according to the actual beam conditions -- the detectors must me movable. For
instance, at early stages of a run, when the beam is not very stable, the
detectors are situated in their home positions. Later, they are inserted into
the beam pipe and placed in the immediate vicinity of the beam.  As the
detectors positions need to be established on the run-by-run basis, one needs a
data-driven method of aligning forward proton detectors. 

\section{Dynamic Alignment Method} 

The dynamic alignment method \cite{Goulianos2010,Gallinaro:2006vz} has been
used by CDF for their RPS (Roman Pot Spectrometer) detector. For a given time
period a sample of single diffractive events was collected. Then four-momentum
transfer distribution was reconstructed assuming different detectors positions.
In particular, one is interested in the distribution value at $t'=0$ (in the
following, primed variables denote the reconstruced values and unprimed
variables are used for the true values of observables):
\[ S = \frac{\d \sigma}{\d t'} \Big|_{t'=0}. \]
When the assumed detectors positions are wrong, such is also the value of the
reconstructed four-momentum transfer. Thus, $S$ is a function of the
misalignment. The dynamic alignment method assumes that $S$ reaches a maximum
value for the perfectly aligned detectors. In the following a simplified
justification of this method is given.

As mentioned before, the detectors measure trajectories of scattered particles,
\textit{i.e.} the trajectory position $(x,y)$ and its elevation angles (slopes)
$(s_x,s_y)$. In the simplest case there are two detector stations per beam,
both measuring proton position: $(x_1, y_1)$ at $z_1$ and $(x_2, y_2)$ at
$z_2$, spaced by the distance $L=z_2-z_1$. The parameters of the trajectory at
the mid point, $z=(z_1+z_2)/2$, are:
\[ x  = \frac{x_1 + x_2}{2}, \quad y  = \frac{y_1 + y_2}{2}, \quad s_x =
\frac{x_2 - x_1}{L}, \quad s_y = \frac{y_2 - y_1}{L}.  \] 

The misalignment of such detectors has four degrees of freedom, for both
stations and both directions: $\Delta x_1$, $\Delta y_1$, $\Delta x_2$, $\Delta
y_2$.  In fact, one should include also possible misalignment of the $z$
detector position and skewness of the coordinate system, but in a real
experimental environment they are negligible. It will prove to be helpful to
use the global:
\[ \Delta x  = \frac{\Delta x_1 + \Delta x_2}{2}, \quad \Delta y  =
\frac{\Delta y_1 + \Delta y_2}{2}, \]
and the relative:
\[ \delta x  = \Delta x_2 - \Delta x_1, \quad \delta y  = \Delta y_2 - \Delta
y_1,  \] 
misalignments, instead of $\Delta x_1, \Delta y_1, \Delta x_2, \Delta y_2$.

A large fraction of protons tagged in forward detectors originate from single
diffractive processes (\textit{i.e.} single diffractive dissociation). The
four-momentum transfer distribution of these events is exponential:
\[ \frac{\d \sigma}{\d t} = \sigma_0 b e^{-bt}, \]
here $\sigma_0$ is the single diffraction cross section and $b$ is the nuclear
slope. When a proton is tagged, its initial momentum (momentum at the
Interaction Point, IP) is unfolded from the measured values of $(x,y,s_x,s_y)$.
In particular: 
\[ p_x = f(x,y,s_x,s_y), \quad p_y = g(x,y,s_x,s_y). \]
Afterwards these values are used to calculate the four-momentum transfer.
However, if the detectors are misaligned, the unfolded momentum differs from
the real one:
\[ p_x' = p_x + \Delta_x, \qquad p_y' = p_y + \Delta_y, \]
where: 
\begin{align*} \Delta_x \approx \ & \frac{\partial f}{\partial x} \Delta x +
\frac{\partial f}{\partial y} \Delta y + \frac{\partial f}{\partial s_x}
\frac{\delta x}{L} + \frac{\partial f}{\partial s_y} \frac{\delta y}{L}, \\
\Delta_y \approx \ & \frac{\partial g}{\partial x} \Delta x + \frac{\partial
g}{\partial y} \Delta y + \frac{\partial g}{\partial s_x} \frac{\delta x}{L} +
\frac{\partial g}{\partial s_y} \frac{\delta y}{L}. \\ \end{align*}

As $t \approx p_x^2 + p_y^2$ and $t' \approx (p_x')^2 + (p_y')^2$,
one can obtain the reconstructed four-momentum transfer distribution (assuming
that $\Delta_x$ and $\Delta_y$ are proton momentum independent):
\[ \frac{\d \sigma}{\d t' \, \d \varphi'} = \frac{\sigma_0 b}{2\pi}
e^{-b\left(t' - 2 \sqrt{t'} \Delta \cos(\varphi'-\alpha) + \Delta^2 \right)},
\]
where $\Delta$, $\varphi'$ and $\alpha$ are defined by:
\begin{equation*}
p_x' = \sqrt{t'} \cos\varphi',\ 
p_y' = \sqrt{t'} \sin\varphi',\ 
\Delta_x = \Delta \cos\alpha,\ 
\Delta_y = \Delta \sin\alpha.
\end{equation*}
Expanding $\exp(2b\sqrt{t'}\Delta\cos(\varphi'-\alpha))$ into a power series
one obtains:
\[ S = \frac{\d \sigma}{\d t'}\Big|_{t'=0} = \sigma_0 b e^{-b\Delta^2}.   \]
This shows that for perfect alignment $S$ indeed reaches its maximal value.

At this point there some important remarks are to be made. Firstly, to be able
to reconstruct the $t$ distribution at $t=0$, such events must be within the
acceptance of the detector. As will be discussed later on it is not always the
case.

Secondly, the derivation presented above assumes that $\Delta_x$ and $\Delta_y$
are constant (not depending on the proton momentum), \textit{i.e.} higher
derivatives of $f$ and $g$ are zero. This assumption is needed only to obtain
the analytic formula for $S$, but is not crucial for the method. However, there
are restrictions on the possible variation of $\Delta_x$ and $\Delta_y$ -- on
the average they must be substantially different from zero, otherwise the net
effect cancels.

Thirdly, the fact that for the perfect alignment $S$ gets maximal does not mean
that by maximisation of $S$ one aligns the detectors. This is because
maximising $S$ is equivalent to requesting $\Delta_x = 0$ and $\Delta_y = 0$.
These two equations do not have a unique solution, as there are four unknown
misalignments. However, when the matrix of partial derivatives of $f$ and $g$:
\begin{equation*} \left( \begin{array}{cc} \partial_x    f  & \partial_x    g
\\ \partial_y    f  & \partial_y    g \\ \partial_{sx} f  & \partial_{sx} g \\
\partial_{sy} f  & \partial_{sy} g \end{array} \right) \end{equation*}
has two rows that are  negligible and the remaining two rows form a matrix with
non-vanishing determinant, the method provides a direct alignment for variables
corresponding to non-negligible rows.  Otherwise one needs another method that
will give the two missing constrains.

Finally, one must remember that there are additional experimental effects that
influence the measurement. Such factors (\textit{e.g.} spatial resolution of
the detectors, beam angular spread) cause random smearing of the
reconstructed four-momentum transfer leading to statistical errors of the
obtained $t$ distribution. Thus, for a given number of collected single
diffractive events there is a limit on the alignment precision.

\section{Alignment at the LHC}

In this section the dynamic alignment method applicability at the LHC is
presented; the forward proton detectors in the ATLAS experiments are considered
-- the ALFA \cite{Franz:2009zz} and the AFP detectors at 220 m (AFP220)
\cite{Royon:2007ah}. 

The ALFA detector stations consist of two roman pots at each outgoing beam,
positioned symmetrically with respect to the IP at 237.4 and 241.5 metres. Each
pot allows to insert vertically the position sensitive and triggering detectors
into the beam pipe. The detectors will use the scintillating fibers to measure
the scattered proton position.

The main purpose of ALFA is to measure the elastic proton-proton scattering in
the Coulomb amplitude dominance region. This allows for precise calculations of
the scattering amplitude and hence precise determination of the luminosity. As
already mentioned, for that purpose a special high $\beta^*$ optics will be
used. This causes that single diffractive events with $t=0$ are not within the
detectors acceptance, for all possible proton momentum losses
\cite{Trzebinski:2010}. Therefore, the dynamic alignment method cannot be used
for the ALFA detectors.

The AFP220 detectors are currently at the R\&D stage and are planned for
installation during the LHC shutdown before the \unit{$10^{34}$}{\lumiunits}
luminosity runs (however, possibilities of earlier installation are also
discussed).

Like for the ALFA experiment, two detector stations per an outgoing beam are
planned -- they will be positioned symmetrically at 226 and 224 metres away
from the ATLAS IP.  However, instead of roman pots, the Hamburg movable beam
pipe mechanism \cite{schneekloth, Albrow:2008pn} will be used to insert the
detectors inside the beam pipe.  Each station will consist of a silicon
detector, for the proton position measurement, and a fast timing detector (with
resolution of several picoseconds) that is necessary for the pile-up background
reduction.

\begin{table}[t]
\caption{Approximate ranges of $f(x,y,x',y')$ and $g(x,y,x',y')$ derivatives
for the standard LHC tune (V6.5, collision \cite{LHCoptics}).}
\[
\begin{array}{c | c c}
                         & f(x,y,x',y')           & g(x,y,x',y') \\ \hline
\partial_x               & (-2 \pm 4) \cdot 10^1  & (-1 \pm 1) \cdot 10^2     \\
\partial_y               & 0                      & 0                         \\
\frac{1}{L}\partial_{sx} & (2.5 \pm 1) \cdot 10^3 & (4 \pm 2.5) \cdot 10^2    \\
\frac{1}{L}\partial_{sy} & 0                      & (-3 \pm 2) \cdot 10^3 \\
\end{array}
\]
\label{tab:derivatives}
\end{table}

As discussed in Section 2, crucial for the method are the derivatives of the
$p_x$ and $p_y$ unfolding functions ($f$ and $g$) with respect to $x$, $y$,
$s_x$ and $s_y$.  Ranges of the derivatives for the AFP220 case are given in
Table~\ref{tab:derivatives}.  These  were calculated for single diffractive
events generated with Pythia 6.4 \cite{Sjostrand:2006za}, transported with
FPTrack \cite{FPTrack} and then unfolded as in \cite{Staszewski:2009sw}.

It can be seen that in the first column the most significant element is the
$\partial_{sx}$ and for the second column the most important is the
$\partial_{sy}$. Therefore, for both columns, the $\partial_x$ and $\partial_y$
rows can be neglected. The remaining $\partial_{sx}$ and $\partial_{sy}$ rows
form a matrix with non-zero determinant, which depends only on the most
significant elements in each column. This shows that the dynamic alignment
method will work for the AFP220 detectors. However, in contrast to the
situation of the CDF detectors, the method is sensitive to the relative
alignment and not to the global one.
 
\begin{figure}[p]
\centering
\includegraphics[width=\linewidth]{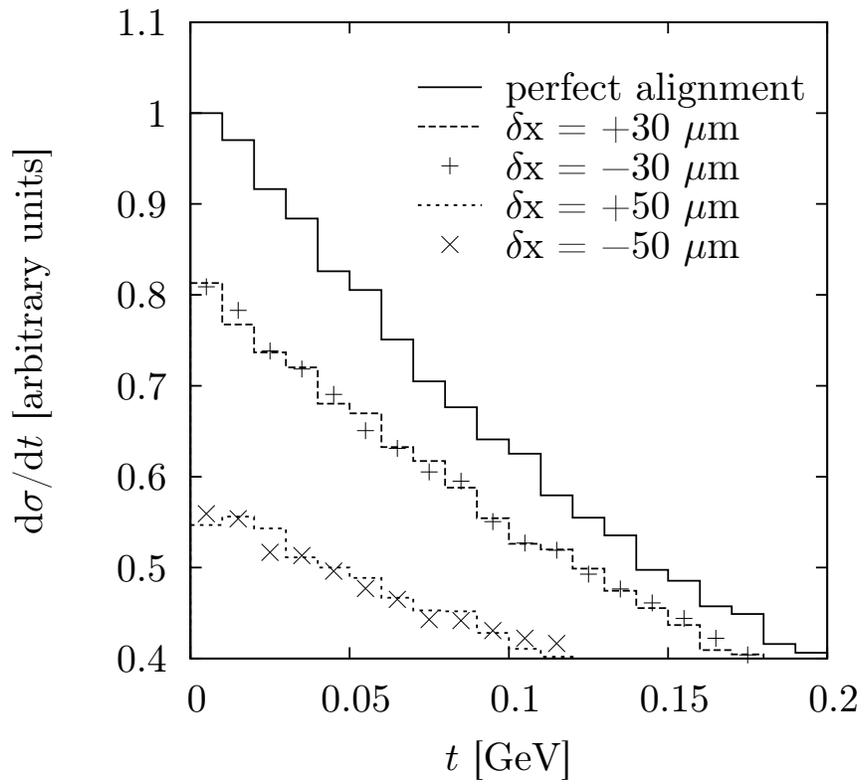}\\
\caption{Reconstructed four-momentum distribution for different values of
horizontal misalignment ($\delta x$).}
\label{fig:dx}
\end{figure}

\begin{figure}[p]
\centering
\includegraphics[width=\linewidth]{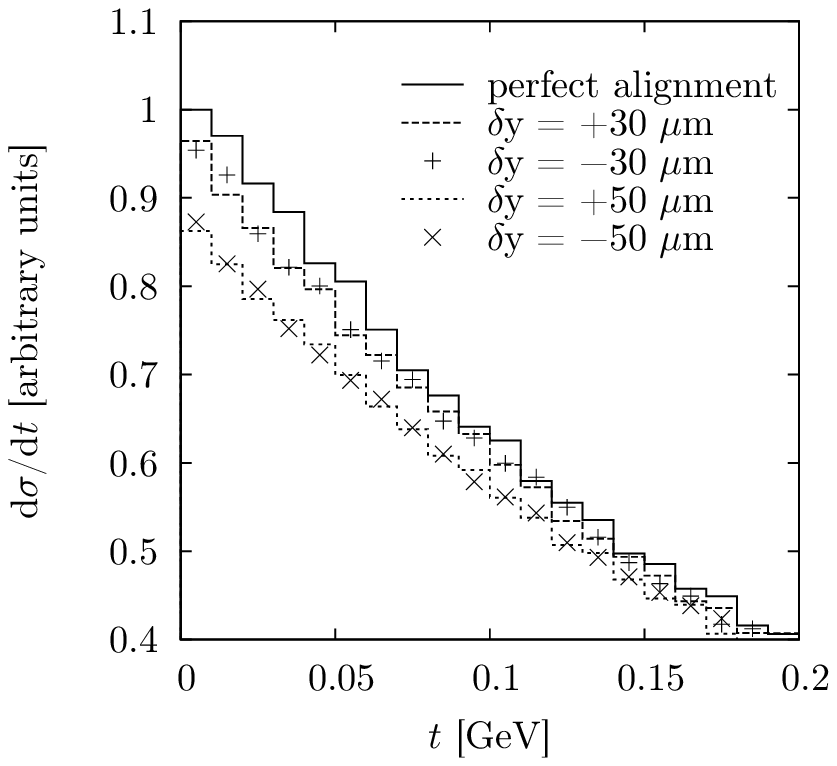}
\caption{Reconstructed four-momentum distribution for different values of
vertical misalignment ($\delta y$).}
\label{fig:dy}
\end{figure}

This is illustrated in Figures \ref{fig:dx} and \ref{fig:dy}, where the
reconstructed four-momentum transfer distributions are presented for different
values of $\delta x$ and $\delta y$: $\unit{0}{\micron}$,
$\pm\unit{30}{\micron}$ and $\pm\unit{50}{\micron}$.  Indeed, the simulation,
which includes the spatial resolution and the multiple Coulomb scattering at
the station at 216 metres \cite{Staszewski:2009sw}, confirms the conclusions
that were obtained from the analysis of the $f$ and $g$ derivatives. The
four-momentum transfer distribution at $t=0$ decreases when relative
misalignment is introduced.  Such  behaviour is not observed in case of the
global misalignment. This follows from  the fact that $\partial_x$ and
$\partial_y$ columns in Table \ref{tab:derivatives} are practically negligible.
Also, as can be seen from Figures \ref{fig:dx} and \ref{fig:dy}, the
sensitivity to the relative misalignment in the vertical direction is smaller
than that for the horizontal one.

\section{Conclusions}

The dynamic alignment method for the forward proton tagging detectors proposed
by the CDF Collaboration was reviewed. A simplified mathematical justification
of this method was given and its applicability at the LHC was discussed. It was
shown that it works there differently than at the Tevaron, where it enabled to
determine the global alignment of the CDF RPS detectors. In the ATLAS
Experiment the method it is not useful for the ALFA detectors, whereas it can
be successfully employed for relative alignment of the AFP220 stations.

\bibliographystyle{utphys.bst}
\bibliography{bibliography}{}

\providecommand{\href}[2]{#2}\begingroup\raggedright\begin{thebibliography}{10}

\bibitem{Albrow:2010yb}
M.~Albrow, T.~Coughlin, and J.~Forshaw, ``{Central Exclusive Particle
  Production at High Energy Hadron Colliders},''
  \href{http://dx.doi.org/10.1016/j.ppnp.2010.06.001}{{\em
  Prog.Part.Nucl.Phys.} {\bfseries 65} (2010) 149--184},
  \href{http://arxiv.org/abs/1006.1289}{{\ttfamily arXiv:1006.1289 [hep-ph]}}.

\bibitem{Staszewski:2009sw}
R.~Staszewski and J.~Chwastowski, ``{Transport Simulation and Diffractive Event
  Reconstruction at the LHC},''
  \href{http://dx.doi.org/10.1016/j.nima.2009.08.023}{{\em Nucl. Instrum.
  Meth.} {\bfseries A609} (2009) 136--141},
\href{http://arxiv.org/abs/0906.2868}{{\ttfamily arXiv:0906.2868
  [physics.ins-det]}}.

\bibitem{Goulianos2010}
K.~Goulianos, ``{The Forward Detectors of CDF and D0},''
\href{http://arxiv.org/abs/1002.3527}{{\ttfamily arXiv:1002.3527 [hep-ph]}}.

\bibitem{Gallinaro:2006vz}
{\bfseries CDF} Collaboration, M.~Gallinaro, ``{Diffractive and exclusive
  measurements at CDF},''
\href{http://arxiv.org/abs/hep-ex/0606024}{{\ttfamily arXiv:hep-ex/0606024}}.

\bibitem{Franz:2009zz}
S.~Franz and P.~Barrillon, ``{ATLAS ALFA-measuring absolute luminosity with
  scintillating fibres},''
  \href{http://dx.doi.org/10.1016/j.nima.2009.05.148}{{\em Nucl.Instrum.Meth.}
  {\bfseries A610} (2009) 35--40}.

\bibitem{Royon:2007ah}
{\bfseries RP220} Collaboration, C.~Royon, ``{Project to install roman pot
  detectors at 220 m in ATLAS},''
\href{http://arxiv.org/abs/0706.1796}{{\ttfamily arXiv:0706.1796
  [physics.ins-det]}}.

\bibitem{Trzebinski:2010}
M.~Trzebinski, J.~Chwastowski, and T.~Sykora, ``{Transport of Protons with High
  Energy Loss Through the LHC Magnetic Lattice for the High $\beta^*$
  Optics}.'' In preparation, 2010.

\bibitem{schneekloth}
U.~Schneekloth, ``{Hamburg beam pipe technical design}.'' {Private
  comunication}, 1994.

\bibitem{Albrow:2008pn}
{\bfseries FP420} Collaboration, M.~Albrow {\em et al.}, ``{The FP420 R\&D
  Project: Higgs and New Physics with forward protons at the LHC},''
  \href{http://dx.doi.org/10.1088/1748-0221/4/10/T10001}{{\em JINST} {\bfseries
  4} (2009) T10001}, \href{http://arxiv.org/abs/arXiv:0806.0302}{{\ttfamily
  arXiv:arXiv:0806.0302 [hep-ex]}}.

\bibitem{LHCoptics}
``{LHC Optics Web Home}.'' {http://cern.ch/lhcoptics}, 2010.

\bibitem{Sjostrand:2006za}
T.~Sjostrand, S.~Mrenna, and P.~Z. Skands, ``{PYTHIA 6.4 Physics and Manual},''
  {\em JHEP} {\bfseries 05} (2006) 026,
\href{http://arxiv.org/abs/hep-ph/0603175}{{\ttfamily arXiv:hep-ph/0603175}}.

\bibitem{FPTrack}
P.~Bussey. {http://ppewww.physics.gla.ac.uk/$\sim$bussey/FPTRACK}, 2007.

\end{thebibliography}\endgroup
\end{document}